\documentclass[twocolumn,prl,aps,showpacs]{revtex4}
\usepackage{graphicx}
\begin{document}
\title{A `` quantum public key '' based cryptographic scheme for 
continuous variables}
\author{Patrick Navez, Alessandra Gatti and Luigi A. Lugiato}
\affiliation{INFM,
Dipartimento di Scienze CC FF MM,
Universita degli Studi dell'Insubria,
Via Valleggio 11,
I-22100 COMO,
Italy}
\date{\today}
\begin{abstract}
By analogy to classical cryptography, 
we develop a "quantum public key" based 
cryptographic scheme in which the two public 
and private keys consist in each of two 
entangled beams of squeezed light.    
An analog message is encrypted   
by modulating the phase of  
the beam sent in public. The knowledge 
of the degree of non classical 
correlation between 
the beam quadratures measured in private and 
in public allows only the receiver to 
decrypt the message.
Finally, in a view towards absolute security, we 
formally prove that any external intervention of 
an eavesdropper makes him vulnerable to any 
subsequent detection. 
\end{abstract}
\maketitle
Quantum correlations (entanglement) between light beams
is a subject of considerable activity leading to
novel applications such as quantum cryptography, quantum
computation and quantum teleportation. As an alternative
to classical cryptography, quantum cryptography methods 
based on correlation of a
single photon pair \cite{Brassard} have been widely studied over the past
years. They have lead to protocoles permitting
to transmit confidential messages safely \cite{Bennett,Lo,Lutkenhaus}.
On the other hand,
few results have been presented for systems with a large
number of photons such as correlated quadratures 
\cite{Reid, Ralph, Hillery}. 
The use of intense photons beams could  
present numerous technological 
avantages in comparison to the use of single photons pulses.
Photon counting detectors are much more 
efficient to detect squeezed light \cite{Reid} and 
its transmission through an optical fiber can be done 
over a much longer distance \cite{Hillery}. 

In particular, Pereira et al. \cite{Pereira} have proposed a scheme in which 
a coherent light beam containing the message is transmitted 
through an adequate superimposition with two entangled beams of 
squeezed light. Since each of the two entangled beams is a noisy 
channel, they need to be recombined in order to 
decrypt the message. In this approach, the noise is used 
to hide the message which becomes thus unreadable 
during the transmission. 

Based on similar considerations, we aim at developping
a ``quantum public key'' based  cryptographic scheme.
The public key method \cite{Brassard,Publickey} 
is widely used nowadays in classical cryptography.
It consists in two keys. The first key is public and
known to everybody i.e. to the sender (Alice) and the 
receiver (Bob)
of the message and a
possible undesirable third party.
On the contrary, the second key is private and known
only by the receiver of the message. The procedure is the
following: Bob sends the public key to Alice 
who uses it  
to encrypt the message; then Alice 
sends back the encrypted information to Bob who
is the only one able to decrypt it thanks to the private key.
In this scheme, the essential idea
is that the encryption is public in the sense that anyone,
including a third party, can encrypt a message but
any decryption process must require necessarily the knowledge
of the second key.
In this manner, messages are transmitted through
public channel with
strict confidentiality.


In this letter, we investigate theoretically
a quantum public key scheme similar to the 
classical one. 
We device a scheme in which 
the ``q-private'' and ``q-public'' keys consist in each 
of the two entangled photon beams respectively. 
We analyse a process to encrypt an analog message 
using the  
``q-public'' light beam. The
confidentiality of the message is guaranteed 
since the second
``q-private'' light beam is needed for the decryption process.
The prefix ``q-'' has been added to remind the 
quantum character of the signal produced which, 
to the contrary of a classical signal, cannot be 
necessarily reproduced. The non-cloning theorem 
indeed prevents making an exact copy of an unknown 
quantum signal. Thus, the q-public beam is 
accessible only to any first user including a
third party. 

The non-cloning theorem is also the basic idea from which
one can prove that the transmission is secure against an eavesdropper 
(Eve) \cite{Bennett, Lo, Lutkenhaus}. 
The impossibility of reproduction prevents 
Eve from having access to any quantum state without modifying it.
Such a modification makes her  
vulnerable to any subsequent detection by the sender and/or 
the receiver. As far as security is concerned,
we formally prove for 
the first time, to our knowledge, the vulnerability of an 
eavesdropper to any external intervention 
during a communication process using continuous variables. 
However, we  
point out that this result does not imply the absolute 
security of the transmission since the vulnerability of Eve 
does not mean that she will be necessarily detected.
An appropriated protocol of communication is required to detect that 
eavesdropping has occured. 
This essential issue is not completely 
discussed in this letter and remains still open in the 
case of continuous variables. 

Let us start with the description of the 
quantum public key based scheme.
Suppose that
Alice wishes to send a message to Bob. 
To this purpose,
Bob produces
an EPR state consisting of two 
entangled beams 
characterized by the photon annihilation 
operators $\hat a_1$ and 
$\hat a_2$ \cite{epr}.  
Written in
occupation photon number representation, this state 
has the form:
\begin{eqnarray}\label{squeeze}
\!\!\!
|\Psi\rangle=
\exp \!\left( r{\hat a_1^\dagger}{\hat a_2^\dagger}-
r {\hat a_1}{\hat a_2} \right)
\!\!|0\rangle_1
|0\rangle_2
=\!\sum_{n=0}^\infty
c_n|n\rangle_1|n\rangle_2
\end{eqnarray}
where $c_n=
(\tanh r )^{n}/\cosh r$ and $r$ is the
squeezing real parameter. 
Bob then sends beam 1 to Alice and keeps beam 2
in his laboratory. 
Since Bob has access to the
beam 2, he is the only one able to carry out any
measurement on the total
wave function (\ref{squeeze}).
Alice or Eve, however, is restricted 
to carry out a measurement of any observable 
concerning the subsystem of the beam 1. 
All the information available to
them is extracted                      
only from the diagonal density matrix resulting 
from the partial
trace of the total wave function 
over the unknown beam 2:
\begin{eqnarray}
{\hat \rho_{1}}=
Tr_{2}(|\Psi\rangle \langle \Psi|)=
\sum_{n=0}^\infty |c_n|^2|n\rangle_1 \,_1 \langle n| 
\end{eqnarray}
With the ``q-public'' beam 
received from Bob, 
Alice encrypts the secret message
by making a unitary transformation $\hat M$
which modifies the total wave function
but not the density matrix. Consequently, $\hat M$ 
should verify the following criteria:
\begin{eqnarray}\label{crit1}
{\hat M}|\Psi \rangle \not= |\Psi \rangle
\\ \label{crit2}
{\hat M}{\hat \rho}_1{\hat M^\dagger} = \hat \rho_1
\end{eqnarray}
Then Alice sends the beam 1 back to Bob which, thanks to the
second ``q-private'' beam is the only one to decrypt the message.
Another alternative is that Alice carries out some 
measurement herself on the beam 1 and sends the results to Bob
through a public classical channel. 
The use of such a classical public 
channel avoids the blockability of the system 
by Eve and thus is secure against a 
split-universe attack \cite{Blow}. 
Since the density matrix has not been alterated in the
encryption process, one  
does not observe the presence of the message 
in the signal received by Bob.

Among the many existing 
possibilities,  
the phase transformation:
\begin{eqnarray}
{\hat M}
=\exp\left(i f(\hat a_1^\dagger \hat a_1)\right)
\end{eqnarray}
satisfies both criteria (\ref{crit1}) and
(\ref{crit2}). $f(x)$ could be any real function. 
For convenience, we 
choose a simple linear function 
$f(x)=\theta x$ which transforms the coefficients
$c_n \rightarrow \exp(i\theta n)c_n$.
The constant $\theta$ is the analog quantity containing 
the message that
Alice encrypts. In comparison with \cite{Pereira}, in 
our encryption process, the message is really ``invisible''  
in the signal and thus cannot be distinguished from the noise. 

After the phase transformation, the
field operator associated to the first beam becomes
$\hat a_{1'}=\hat M \hat a_1 \hat M^\dagger=
\exp(-i\theta)\hat a_1$. 
An efficient way to decrypt the information is
to measure observables for which the EPR state
is an eigenstate. 
Because of the difficulty to get access to 
some of them 
in experiment, one uses rather the
quadrature components operators:
\begin{eqnarray}
\hat Z_{1'} &=& \hat a_{1'}e^{-i\phi_A}+\hat a_{1'}^\dagger e^{i\phi_A} =
e^{-i\theta_A}\hat a_1+e^{i\theta_A}\hat a_1^\dagger
\\
\hat Z_{2}&=&
e^{-i\theta_B}\hat a_2+e^{i\theta_B}\hat a_2^\dagger \; ,
\end{eqnarray}
where 
$\theta_A= \phi_A+\theta$. $\phi_A$ and  
$\theta_B$ are the phases of the local 
oscillators used by Alice and Bob, respectively, 
in their homodyne measurements
and determine which quadrature they select for their measurements.
$\theta_B$ must remain private to Bob 
whereas $\phi_A$ must be communicated 
in public to Bob at some stage of the protocol.

Although the EPR state is not
an eigenstate of these operators,  
the uncertainty in the quadrature difference 
$\hat Z_-=\hat Z_{1'}-\hat Z_{2}$ is close to zero 
as the squeezing
parameter $r$ becomes large and 
$\theta_A +\theta_B =0$.
We notice indeed from the expression (\ref{squeeze}) that:
\begin{eqnarray}
\langle \Psi |\delta^2 \hat Z_-|\Psi \rangle
&=& 2\left[ \cosh 2r-\cos(\theta_A+\theta_B) \sinh 2r\right]
\\
&\stackrel{r>>1} {\to}&
4\sinh 2r \sin^2{(\theta_A +\theta_B) \over 2}  
\end{eqnarray}
On the other hand, the introduction of 
non opposite phases $\theta_A+\theta_B \ne 0$
generates for large $r$ 
a quantum uncertainty which appears under the 
form of fluctuations during the 
measurement of the quadrature difference. In this 
manner,
the message is obtained by determining 
the intensity of the noise resulting from these 
fluctuations \cite{epr}.

\begin{widetext}
\begin{figure}[here]
\scalebox{0.6}{
\includegraphics{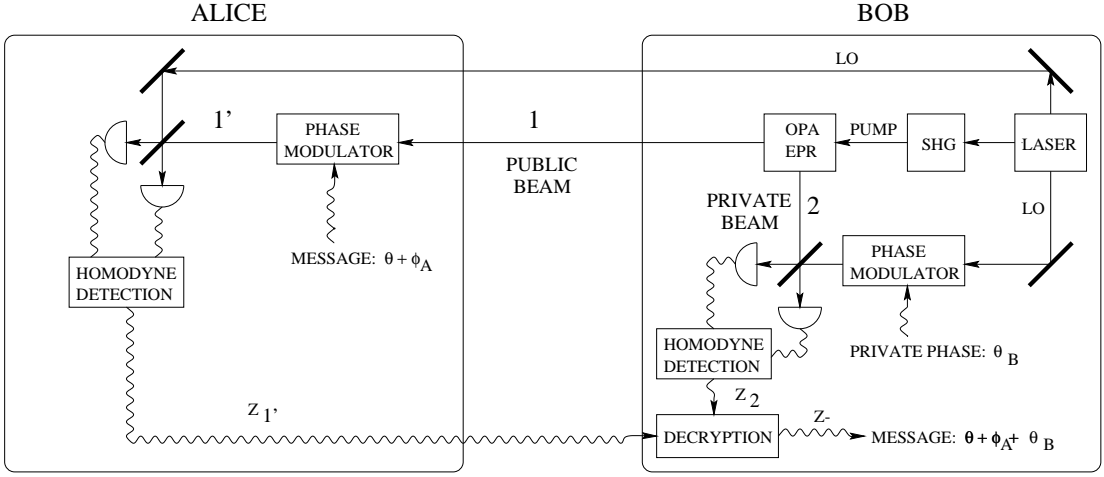}}
\caption{Schematic set-up for the
``quantum public key'' based cryptography with
continous variables}
\label{fig}
\end{figure}
\end{widetext}

Fig.\ref{fig} depicts a possible setup. The two EPR  
``q-private'' 
and ``q-public'' beams are generated through a 
nondegenerate  
parametric down conversion process. They 
result from the 
fluorescence of a pump beam passing through 
a type II crystal acting also as an optical parametric 
amplifier (OPA) \cite{Pereira,Kimble,Yariv,epr}.
The quadratures components are measured 
in a homodyne detection with the help of  
local oscillators fields (LO). 
The pump field, which generates the 
two EPR beams via parametric down 
conversion, is obtained by one intermediate 
second harmonic (SHG) step as usually. 
While Bob
carries out the homodyne 
detection of one quadrature of beam 2 
for phase $\theta_B$, 
Alice encrypts the message $\theta$ 
by modulating the beam 1 phase (e.g. by means
of an electro-optical modulator), before carrying out the measurement 
of the quadrature $\phi_A$. 
Finally, Alice sends back to Bob the 
results of her measurement through a public classical channel. 

Let us now discuss the matter of the vulnerability
of Eve to any subsequent detection by Alice and/or Bob.
Let $\hat Z_{1'}$ and $\hat Z_{2}$ be the 
observables measured  
by Alice and Bob respectively.
Suppose that Eve tries to have access to 
the ``q-public'' beam 1 and thus modifies it 
by means of an unitary 
transformation:
\begin{eqnarray}
\hat a_{1E} =
\hat U^\dagger \hat a_1 \hat U
\end{eqnarray}
The unitary operator $\hat U$ could 
depend on other external degrees of freedom (observables) introduced 
by Eve.  For example, Eve could
use another photon beam or other modes of the same 
beam but taken at a different time or even a detector 
in view of 
a measurement on the beam 1. We denote 
by $|\nu \rangle$ a basis of the Hilbert space characterizing 
these external degrees of freedom and assume that 
$|0\rangle$ is the initial state before the modification.
Then the state after the unitary transformation is 
$\hat U |\Psi\rangle |0 \rangle$.  
Let us define the
probability distribution to find the system with the values 
$z_1$ and $z_2$ of the
quadrature component observables $\hat Z_{1'}$ and
$\hat Z_2$:
\begin{eqnarray}
P(z_{1'},z_2)=\langle \Psi | \delta(z_{1'}-\hat Z_{1'})
\delta(z_2 - \hat Z_2)| \Psi \rangle
\end{eqnarray}

After Eve's action the
probability distribution becomes:
\begin{equation}
\label{PE}
P_E(z_{1'},z_2)=\langle 0 |\langle \Psi |
\hat U^\dagger \delta(z_{1'}-\hat Z_{1'})
\delta(z_2 - \hat Z_2)
\hat U| \Psi \rangle |0 \rangle
\end{equation}
To avoid that Eve presence is revealed, both probability 
distributions must be equal for the 
particular values of $\theta_A$ and 
$\theta_B$ chosen by Alice and Bob. 
On the other hand, Eve is vulnerable to detection,
whenever  
there exist possible choices
of $\theta_A,\theta_B$ (possible homodyne measurements made by Bob or Alice)
for which the probability distribution of the measurement outcomes 
is changed by Eve's action.

{\bf THEOREM :} 
If Alice measures only one quadrature 
$\hat Z_{1'}$, 
Eve attack is not vulnerable to a 
successive detection by Bob if and only if
$\langle \nu |\hat U|0 \rangle$ 
is a function  
only of the quadrature operator $\hat Z_{1'}$. 

PROOF: 
When $\langle \nu |\hat U|0 \rangle$ commutes with 
$\hat Z_{1'}$ then 
the $\hat U$'s cancel each other in 
(\ref{PE}) and trivially:
\begin{eqnarray}\label{probeq}
P_E(z_{1'},z_2)=P(z_{1'},z_2) 
\end{eqnarray}
On the contrary,
let us examine the consequence of requiring
(\ref{probeq})
for all possible choices of the quadrature $\hat Z_2$
measured by Bob but only for the state $|\Psi\rangle$
given by (\ref{squeeze}).
Taking the Fourier transform on $z_1$ and $z_2$, the 
equation (\ref{probeq}) becomes:
\begin{equation}\label{probeq1}
\langle 0 |\langle \Psi |
\hat U^\dagger e^{i\hat Z_{1'}s_1} e^{i\hat Z_2 s_2} 
\hat U| \Psi \rangle |0 \rangle
=
\langle \Psi |
e^{i\hat Z_{1'}s_1} e^{i\hat Z_2 s_2}
| \Psi \rangle
\end{equation}
This equality should be satisfied for any 
value of the real parameters $s_1$, $s_2$, 
$\theta_B$. The last two can be replaced by the more 
global complex parameter
$\xi= \xi_X + i \xi_Y= e^{i\theta_B} s_2$, 
in such a way that 
$\hat Z_2 s_2= \xi \hat a_2^\dagger + \xi^\star \hat a_2$.
Let us introduce:
\begin{eqnarray}
T_{n,n'}(\xi,\xi^*)=
\,_2\langle n'|
e^{-i \left(\xi \hat a_2^\dagger + \xi^\star \hat a_2\right)}
| n \rangle_2
\end{eqnarray}
A calculation shows that for 
all occupation number $n$ and $n'$ of the beam 2:
\begin{eqnarray}\label{lt}
\int{d^2\xi \over \pi}
T_{n,n'}(\xi,\xi^*)e^{ i\left(\xi \hat a_2^\dagger + \xi^\star \hat a_2\right) } =|n\rangle_2 \,_2\langle n'| 
\end{eqnarray}
Applying this transformation to (\ref{probeq1}) and using 
the property of entanglement of 
(\ref{squeeze}), we eliminate 
the states describing the second beam since $\hat U$ does not 
affect beam 2. We obtain for 
all occupation numbers $n$ and $n'$ for beam 1:
\begin{eqnarray}\label{probeq3}
\langle 0 |\,_1\langle n' |
\hat U^\dagger e^{i\hat Z_{1'}s_1} 
\hat U| n \rangle_1|0 \rangle
=
\,_1\langle n' |
e^{i\hat Z_{1'}s_1}
|n \rangle_1
\end{eqnarray}
Operating the inverse transformation of the matrix element in the right
hand side, we get:
\begin{eqnarray}\label{probeq4}
\langle 0 |\,_1\langle n' |
\hat U^\dagger e^{i\hat Z_{1'}s_1}              
\hat U e^{-i\hat Z_{1'}s_1}| n \rangle_1 |0 \rangle
=\delta_{n',n}
\end{eqnarray}
Without loss of generality we can write the explicit dependance 
of $\hat U$ on the 
observables associated to the first beam as $\hat U = 
\hat U(\hat Z_{1'},\hat Q_{1'})$, where 
$\hat Q_{1'}= ( \hat a_1 e^{-i \theta_A} - \hat a_1^\dagger e^{i \theta_A}) /i$ is the 
observable canonically conjugated to $\hat Z_{1'}$.
Noticing that the exponential operator  
is the generator of translations 
i.e.  
$e^{i\hat Z_{1'}s_1} \hat Q_{1'} e^{-i\hat Z_{1'}s_1}= \hat Q_{1'} -2s_1$,
Eq.(\ref{probeq4}) becomes:
\begin{equation}\label{probeq5}
\langle 0 |\,_1\langle n' | 
\hat U^\dagger (\hat Z_{1'},\hat Q_{1'})
\hat U (\hat Z_{1'},\hat Q_{1'}-2s_1)
| n \rangle_1 |0 \rangle
=\delta_{n',n}
\end{equation}
Because two normalized states with unity scalar product 
are necessarily equal, we infer 
that for all $n$:
\begin{eqnarray}\label{probeq6}
\hat U (\hat Z_{1'},\hat Q_{1'})
| n \rangle_1 |0 \rangle=
\hat U (\hat Z_{1'},\hat Q_{1'}-2s_1)
| n \rangle_1 |0 \rangle
\end{eqnarray}
We conclude that, for any state $ |\nu \rangle$,  
$\langle \nu | \hat U (\hat Z_{1'},\hat Q_{1'})|0 \rangle$
does not depend on 
$\hat Q_{1'}$  or equivalently 
it depends only on the quadrature operator $\hat Z_{1'}$ measured 
by Alice. 

{\bf COROLLARY:}
If Alice chooses to measure between at least two distinct 
and non opposite quadratures, 
Eve attack is not vulnerable if and only if $\hat U$
does not interact with
the beam 1 and acts only onto
the state $|0 \rangle$ i.e.:
\begin{eqnarray}\label{theo}
\langle \nu| \,_1\langle n'|\hat U |n\rangle_1 |0 \rangle
=\delta_{n',n} u_\nu
\end{eqnarray}
where $u_\nu$ 
does not depend on $n$.

PROOF: 
From the theorem, we deduce that 
for each non opposite quadrature we must 
have the dependance $\hat U (\hat Z_{1'})$. 
Since the quadratures are independent, the 
only possibility is that 
$\hat U$ is independent of any observable 
relative to the beam 1 or is of the form 
(\ref{theo}).

One direct consequence of the theorem 
and its corollary is 
that if there is 
any element of 
randomness in the choice of the
quadrature $\hat Z_{1'}$ made by Alice, then 
Eve cannot safely extract any information about 
beam 1 and therefore about the message. 
This element of randomness can be introduced as a random choice between
two distinct and non opposite values $\phi_A$. 
Alternatively the message 
itself can be a random sequence of phase shifts $\theta$, chosen at
least between two distinct and non opposite values. 
Thus, Eve is also vulnerable in the case 
of a digital 
transmission in which for example, Alice restricts her measurement to two 
orthogonal quadratures. 

The facts that the probability distributions 
must be equal for any value of $\theta_B$, 
that the second beam has been kept private and 
that $|\Psi\rangle$ has the  
entangled form (\ref{squeeze})
permits to 
achieve the proofs. If the value 
of $\theta_B$ remains fixed, the linear transformation 
cannot be carried out since the integration in (\ref{lt}) must
be done over all $\xi$.
If Eve has 
access to the second beam, then the unitary operator 
depends also on $\hat a_2$ and $\hat a_2^\dagger$ and 
the passage from (\ref{probeq1}) to (\ref{probeq3}) 
is not necessarily valid. Finally, if 
$|\Psi\rangle$ is different and for example 
is a disentangled state, then the beam 1 
might be cloned by Eve and therefore she can 
obtain information without being detected.

The vulnerability of Eve does not mean that 
she will be necessarily detected by Bob. 
When Bob 
receives a message, how can he know 
{\it a priori} that 
eavesdropping has occured? For example,
Eve can block Bob's public beam 
and can replace it with a public beam of her own.  Alice proceeds 
as planned and sends the results of her 
measurement to both Bob and to Eve. 
Eve decrypts the correct message but Bob gets the wrong 
message. To avoid this situation,
Bob must be able to 
check that some expected correlations 
are recovered. 
One possibility is the introduction of 
some redundancies in the 
message which can help him to detect the presence of 
the 
third party. For example, 
Alice sends twice 
portions of the message. Any external intervention 
is detected if two identical portions are not recovered. 
Therefore, a strategy of communication or protocol is 
needed in order to guarantee the absolute security of the 
transmission. The study of such 
protocols requires further investigation and is beyond 
the scope of this letter. 

In summary, we developed  a ``quantum public key''
scheme for encrypting a message by means
of quantum correlated beams.
This scheme is based on the principle that 
any decryption of the message requires both  
measurements of a ``q-private'' key signal and an encrypted
``q-public'' key signal.
Directions for a future research work include, in addition to 
the study of secure protocols, the imperfection in the 
photon counting of the 
detector and the possibility of loss in the transmission
which degrades the correlations.

{\bf ACKNOWLEDGEMENTS}

The authors warmly thank
S. M. Barnett for useful remarks and criticisms.
This work was supported by the network QSTRUCT of the 
TMR programme of the EU.

\end{document}